\documentclass[]{pasj01}
\usepackage{ulem}
\draft

\begin{document} 
\Received{2017/1/28}%{yyyy/mm/dd}
\Accepted{2018/1/13}%{yyyy/mm/dd}
%\Published{yyyy/mm/dd}

\title{Molecular gas in the H{\sc ii}-region complex RCW 166; possible evidence for an early phase of cloud-cloud collision prior to the bubble formation}

%%% begin:list of authors
% Do NOT capitalize all letters in "textsc".
\author{Akio \textsc{Ohama}\altaffilmark{1}, Mikito \textsc{Kohno}\altaffilmark{1}, Shinji \textsc{Fujita}\altaffilmark{1}, Daichi \textsc{Tsutsumi}\altaffilmark{1}, Yusuke \textsc{Hattori}\altaffilmark{1}, Kazufumi \textsc{Torii}\altaffilmark{2}, Atsushi \textsc{Nishimura}\altaffilmark{1}, Hidetoshi \textsc{Sano}\altaffilmark{1}, Hiroaki \textsc{Yamamoto}\altaffilmark{1}, Kengo \textsc{Tachihara}\altaffilmark{1} and Yasuo \textsc{Fukui}\altaffilmark{1}}
\altaffiltext{1}{Department of Physics, Nagoya University, Furo-cho Chikusa-ku Nagoya, 464-8602, Japan}
\altaffiltext{2}{Nobeyama Radio Observatory, 462-2 Nobeyama, Minamimaki Minamisaku, Nagano, 384-1305, Japan}
\email{ohama@a.phys.nagoya-u.ac.jp}

%%% end:list of authors

%% `\KeyWords{}' always has to be placed before `\maketitle'.
 %Do NOT move this preamble from here!

\maketitle

\begin{abstract}
Young H{\sc ii} regions are an important site to study O star formation based on distributions of ionized and molecular gas. 
We revealed that two molecular clouds at $\sim 48$ km s$^{-1}$ and $\sim 53$ km s$^{-1}$ are associated with the H{\sc ii} regions G018.149-00.283 in RCW 166 by using the JCMT CO High-Resolution Survey (COHRS) of the $^{12}$CO ($J$=3--2) emission.  
G018.149-00.283 comprises a bright ring at 8 $\mu$m and an extended H{\sc ii} region inside the ring. 
The $\sim 48$ km s$^{-1}$ cloud delineates the ring, and the $\sim 53$ km s$^{-1}$ cloud is located within the ring, indicating a complementary distribution between the two molecular components.
We propose a hypothesis that high-mass stars within G018.149-00.283 were formed by triggering in cloud-cloud collision at a projected velocity separation of $\sim 5$ km s$^{-1}$. 
We argue that G018.149-00.283 is in an early evolutionary stage, $\sim 0.1$ Myr after the collision according to the scheme by \citet{hab92} which will be followed by a bubble formation stage like RCW 120. 
We also suggested that nearby H{\sc ii} regions N21 and N22 are candidates for bubbles possibly formed by cloud-cloud collision. 
\citet{ino13} showed that the interface gas becomes highly turbulent and realizes a high-mass accretion rate of $10^{-3}$ -- $10^{-4}$ $M_{\odot}$ $/$yr by magnetohydrodynamical numerical simulations, which offers an explanation of the O-star formation. 
%As a byproduct of the study, we identified another case of cloud-cloud collision toward G018.305-00.391 which is probably located at 13.3 kpc on the far side of the galactic disk.
A fairly high frequency of cloud-cloud collision in RCW 166 is probably due to the high cloud density in this part of the Scutum arm.

\end{abstract}

\section{Introduction}
The feedback effects by high-mass stars including UV radiation, strong stellar winds, super nova explosions in the end of their lives exert profound effects on galactic evolution and star formation. It is therefore an important issue in astrophysics to understand the mechanism of high-mass star formation. In spite of tremendous efforts devoted to resolve the issue, both in theoretical and observational points of view, details of high-mass star formation is deeply veiled \citep{zin07}.

\citet{hab92} developed numerical simulations on supersonic collision between uneven clouds and the study was followed-up by \citet{ana10} and \citet{tak14}; these authors made numerical simulations of collision between a small cloud and a large cloud and showed that a strongly collision-compressed interface provides physical conditions with much higher density than prior to the collision favorable for enhanced star formation. The morphology of the collision is characterized by a cavity in the large cloud created by the small cloud, where stars are formed by triggering. \citet{ino13} made magnetohydrodynamical (MHD) numerical simulations of the collisional interface and showed that massive dense cloud cores are formed in the compressed layer. Their simulations demonstrated that the supersonic collision creates turbulent gas motion with amplified magnetic field leading to a high-mass accretion rate of 10$^{-4}$ -- 10$^{-3}$ $M_{\odot}$ yr$^{-1}$. Such a mass accretion rate satisfies what is required to form high-mass stars by overcoming the stellar feedback of high radiation pressure \citep{wol87,mck03}.

The ultraviolet radiation emitted by high-mass stars ionizes the ambient neutral gas and creates H{\sc ii}  regions of various sizes as listed in a number of catalogs of H{\sc ii}  regions (e.g., Sharpless 1959; Rodgers et al 1960; Churchwell et al. 2006; 2007).
 In the early phase of high-mass star formation, high-mass stars are located in dense molecular cloud cores and they produce compact and dense H{\sc ii}  regions. These are called ultra-compact H{\sc ii} regions and possibly  an early phase of H{\sc ii}  region evolution. They may evolve into more extended H{\sc ii}  region later like compact H{\sc ii}  regions or usual H{\sc ii}  regions \citep{chu06}. The previous studies observed molecular clouds in extended H{\sc ii}  regions with a size of 10 pc and found evidence for cloud-cloud collision which triggered formation of superstar clusters; Westerlund2 \citep{fur09,oha10}, NGC3603 \citep{fuk14} and 
  RCW38 \citep{fuk16}. The super star clusters are a dense stellar system including stars of 10$^4$ $M_{\odot}$ in $\lesssim$ 1 pc$^3$. A common property among these molecular clouds is that the molecular column density is as high as 10$^{22-23}$ cm$^{-2}$ and the relative velocity between the colliding clouds is 10 - 20 km s$^{-1}$. This large velocity separation is not explained as due to stellar winds or supernova explosions, but is due to large scale gas acceleration by collective stellar feedback like super shells, or galactic spiral arms/stellar gravity over $\geq$ 1000 Myrs. The observations suggest that such a supersonic collision between two clouds causes trigger of super star clusters or O star (s) in 0.1 - 1 Myr, a crossing time scale of a 1 - 10 pc cloud by $\sim$ a relative velocity 10 km s$^{-1}$.

Detailed observational investigations of H{\sc ii}  regions having a size of 1 - 10 pc discovered isolate O star formation by collision between molecular clouds of 10$^3$ $M_{\odot}$; M20 \citep{tor11,tor17} and RCW 120 \citep{tor15}). Common properties of these clouds are that the two clouds show complementary distribution and bridge features linking them in velocity. \citet{tor17} argued that synthetic observations of colliding clouds of numerical simulations by \citet{tak14} reproduce the bridge features in M20. In addition, Haworth et al.(2015ab) showed that bridge features are consistent with the simulations and presented constraints on the lifetime of the bridge features. Recently, cloud-cloud collision was reported in the nearest high-mass star forming regions M42, NGC 2024 and NGC 2068/2071 in Orion \citep{fuk18,oha17,tsu18}. %(Fukui et al. 2017; Ohama et al. 2017; Tsutsumi et al. 2017)
 These authors proposed that cloud-cloud collision triggers O star formation in seemingly single-velocity clouds, whereas they comprise always two velocity components as revealed by detailed analyses of the cloud velocity field. 

Figure 1a shows three color images of the RCW 166 region at mid-infrared radiation with Spitzer (blue  3.6 $\mu$m, green 8 $\mu$m, and red 24 $\mu$m) \citep{chu06}. 
The 3.6 $\mu$m represents stars, the 8 $\mu$m mainly the PAHs, and the 24 $\mu$m warm dust in H{\sc ii} regions. 
RCW 166 comprises two Spitzer bubbles N21 and N22 and an H{\sc ii} region G018.149-00.283, which are located in a giant molecular cloud of $\sim$ 10$^5$ $M_{\odot}$ as shown by \citet{par13}. 
In addition, supernova remnant (SNR) G18.1-0.1 is located on the west side of G018.149-00.283, although it is not clear if the SNR is interacting with the molecular cloud.
\citet{loc89} detected radio recombination lines in N21 at 43.2 km s$^{-1}$. 
\citet{par13} showed that one O star is located in N 21 (shown by a triangle in Figure 1) based on a near infrared and spectroscopic optical study.
\citet{wat08} identified 21 YSO candidates around N21. 
N22, including 11 O-star candidates \citep{ji12}, might shows more active star formation than N21.
\citet{kol03} detected radio recombination line in N22 at 51 km s$^{-1}$. \citet{and09a} estimated distances of N21 and N22 to be 3.6 kpc and 4.0 kpc from H{\sc i} absorption \citep{mcc07} by assuming the Galactic rotation curve \citep{bra93}. 
\citet{par13} also identified an early O star toward a nearby H{\sc ii} region (G018.237-00.240) from spectroscopic observations obtained with the 2.15-m telescope at CASLEO. 
Toward a nearby compact source G018.305-00.391 \citet{loc89} detected radio recombination lines at 32.7 km s$^{-1}$, and \citet{wie12} detected NH$_3$ at 32.3 km s$^{-1}$ with Green Bank Telescope.
The relation between G018.305-00.391 and RCW 166 is unclear.

\citet{and09a} analyzed molecular clouds in 301 H{\sc ii}  regions by using the dataset Galactic Ring Survey (GRS) taken by Boston University and Five College Radio Astronomy Observatory. \citet{par13} made a detailed analysis of GRS and identified molecular gas associated with N21, N22, and G018.149-00.283. These authors estimated the molecular mass directly associated with RCW 166 to be 1 $\times$ 10$^5$ $M_{\odot}$ by assuming the local thermodynamic equilibrium from the integrated intensity of the $^{13}$CO ($J$=1--0) emission. \citet{bro96} detected CS ($J$=2-1) emission at 54.9 km s$^{-1}$ in RCW 166. In order to obtain data at higher angular resolution than GRS we used the $^{12}$CO ($J$=3--2) in CO High-Resolution Survey (COHRS) with JCMT \citep{dem13} and focused ourselves on detailed molecular distribution and kinematics in the central 20 pc of RCW 166, where the molecular gas is densest and star formation is most active. Section 2 describes the datasets, Section 3 gives results, Section 4 develops discussion on cloud-cloud collision, and Section 5 presents conclusions. 

\section{Datasets}
The present work used three datasets of GLIMPSE I, MIPSGAL, and COHRS. GLIMPSE I \citep{ben03} presents infrared images of the inner Galaxy taken with IRAC \citep{faz04} on board Spitzer Space Telescope \citep{wer04} for 10$^{\circ}$ $\textless$  $l$ $\textless$ 65$^{\circ}$ and 295$^{\circ}$ $\textless$ $l$  $\textless$  350$^{\circ}$, and -1$^{\circ}$ $\textless$ $b$ $\textless$ 1$^{\circ}$. The angular resolutions are from 1.5 arcsec (3.6 $\mu$m) to 1.9 arcsec (8.0 $\mu$m) at wavelengths 3.6, 4.4, 5.8, and 8.0 $\mu$m. MIPSGAL \citep{car05} used Multiband Imaging Photometer for Spitzer (MIPS) on board Spitzer Space Telescope at 24 $\mu$m and 70 $\mu$m (Version 3.0) for 292 $^{\circ}$ $\textless$ $l$ $\textless$ 69 $^{\circ}$ and -1$^{\circ}$ $\textless$ $b$ $\textless$ 1$^{\circ}$. Angular resolutions are 5 arcsec at 24 $\mu$m and 15 arcsec at 70 $\mu$m. 

In COHRS \citep{dem13} observations of the $^{12}$CO($J$=3--2) emission were made with Heterodyne Array Receiver Program (HARP; \citet{buc09}) installed at James Clerk Maxwell Telescope (JCMT). These observations covered 29 square degrees listed in Table 1 of \citet{dem13}. HARP is a superconductor-insulator-superconductor (SIS) receiver of 16 (4$\times$4) pixels separated by 30 arcsec. The beam size and the main-beam efficiency are 14 arcsec and 0.61, respectively. The spectrometer has a frequency resolution of 0.488 MHz or a velocity resolution of 0.42 km s$^{-1}$ with a 1 GHz coverage. The archival data covers -30 km s$^{-1}$ $\textless$ V$_{lsr}$ $\textless$ 155 km s$^{-1}$. The observations were made in the on-the-fly mode and a grid spacing is taken to be 7.3 arcsec about a half of the main beam size. Pointing accuracy is about 2.0 - 2.5 arcsec. Typical rms noise is about 2 K ch$^{-1}$ in T*a for 0.42 km s$^{-1}$ resolution.

\section{Results}
\subsection{Global molecular gas distribution in RCW 166}
Figure 1b shows integrated intensity distribution of $^{12}$CO $J$=3--2 integrated from 39.6 to 54.6 km s$^{-1}$, when we see the CO features toward N21 and N22.
Figure 2 shows a velocity channel distribution of the $^{12}$CO $J$=3--2 emission for a velocity range from 43.6 km s$^{-1}$ to 55.6 km s$^{-1}$ where molecular clouds on the near side of the Scutum arm \citep{rei16} are seen.
Molecular clouds in $> \ \sim$60 km s$^{-1}$ are probably in Norma arm \citep{rei16} and not interacting with N21, N22, G018.305-00.391, and G018.149-00.283, and hence we do not treat the clouds in this paper.
By focusing on the H{\sc ii} regions and multiple O stars, signatures of active formation of young high-mass stars, we recognized the following three regions of distinct star formation in Figure 1a as already described and discussed at a lower angular resolution of GRS by \citet{par13}; one is the main CO peak toward the H{\sc ii} region G018.149-00.283 (18.15$^{\circ}$, -0.3$^{\circ}$), another N21 at (18.18$^{\circ}$, -0.45$^{\circ}$ -- -0.4$^{\circ}$), and the other N22 at (18.25$^{\circ}$ -- 18.30$^{\circ}$, -0.32$^{\circ}$ -- -0.28$^{\circ}$). 
%We also found the fourth region of rather isolated single O star formation at (18.25$^{\circ}$, -0.25$^{\circ}$). 
Parameters of each region are summarized in Table 1.

\subsection{Molecular gas toward N21 and N22}
Figure 3 shows the $^{12}$CO($J$=3--2) distributions toward N21 and N22. In panels b and c, longitude-velocity diagrams for N21 and N22, we found no significant line broadening toward N21 and N22 and suggest that the two show no signatures of expansion with velocity greater than $\sim$ 3 km s$^{-1}$.
Figure 4 picks up three velocity ranges [blue-shift 43.6 - 47.6 km s$^{-1}$, intermediate 47.6 - 50.6 km s$^{-1}$, and red-shift 51.6 - 55.6 km s$^{-1}$] for a closer look, which are overlaid on the Spitzer infrared image. 
By comparing the CO distribution with the bright infrared emission, we found the three regions to be of particular interest in connection with the young high-mass stars; 
%1) G018.149-00.283 is associated with strongest CO emission in a broad velocity range over $\sim$ 20 km s$^{-1}$ in every panel of Figure 4, 
1) in Figure 4a N21 is associated with an elliptical ring-like CO feature of $\sim$ 3 pc diameter, which coincides with the infrared ring, 
and appears to be associated with a filamentary CO feature inside the ring at $(l, b) = (18.12^{\circ}) - 18.24^{\circ}, -0.45^{\circ} - -0.36^{\circ})$ and 
2) in Figure 4b N22 is associated with an intermediate velocity ring-like CO feature showing a correspondence with the infrared ring and includes 11 O-star candidates.

\subsection{Molecular gas toward G018.149-00.283}
Figure 5 shows details of the molecular gas toward the H{\sc ii} region G018.149-00.283 for the two velocity ranges, the blue-shifted cloud 47.6 - 49.6 km s$^{-1}$ (5a) and the red-shifted cloud 51.6 - 55.6 km s$^{-1}$ (5b). 
The molecular gas in the intermediate velocity range 47.6 -- 49.6 km s$^{-1}$ (intermediate cloud) is shown by black contours. Three squares represent the position of B stars identified by \citet{par13}. 
Figure 5c shows an overlay of the two components, and indicates that they show complementary distribution. 
%The young objects, the H{\sc ii} region and the O-star candidate, are located in the red-shifted cloud. 
We introduced another coordinate (R, S) where R-axis is taken to be perpendicular to the elongation of the red-shifted cloud (position angle is $45^{\circ}$ in Figure 5), and S-axis orthogonal to R-axis.
The origin of (R, S) is forward the center of the H{\sc ii} region $(l, b) = (18.15^{\circ}, -0.28^{\circ})$. 
%移動
%\citet{tak14} made hydrodynamical numerical simulations of two colliding spherical clouds. One of the clouds, the small cloud, has a radius of 3.5 pc an the other, the large cloud has a radius of 7.2 pc. Details of the head-on collision are summarized in Table 2. Figure 7a shows a schematic view of the collision seen from the direction perpendicular to the cloud relative motion at two epochs, 0.0 Myr and 1.6 Myr, after the onset of the collision based on the numerical simulations. The small cloud is producing a cavity in the large cloud by the collisional interaction at 1.6 Myr. The interface layer of the two clouds has enhanced density by collision, and the internal turbulence mixes the inhomogeneous gas of the clouds into the layer. Figures 7b-7i show velocity-channel distributions every 0.9 km s$^{-1}$ as seen from a direction $\theta$ = $45^{\circ}$ tilted to the relative motion.
%The small cloud is seen at a velocity range from 2.0 to 4.7 km s$^{-1}$, and the large cloud from -1.5 to 1.1 km s$^{-1}$.
%The velocity range from 1.1 to 2.0 km s$^{-1}$  corresponds to the intermediate velocity layer.
%Note that the velocity ranges of each panel in Figures 7b-7i do not exactly fit the velocity ranges of the two clouds in part due to the internal turbulent.
%The small cloud is flattened perpendicular to the traveling direction by the merging into the intermediate layer, and the large cloud shows intensity depression corresponding to the cavity created by the small cloud.
Figures 7a and 7b show close-ups of Figures 5a and 5b with (R, S) coordinate.
Figure 7c shows a position-velocity cut nearly perpendicular to the elongation of the red-shifted cloud.
The figure shows that the cloud is continuous over 8 km s$^{-1}$ at a velocity gradient of 2 km s$^{-1}$/pc in the direction perpendicular to the ring.
%移動
%Figure 8d shows simulations of the cloud velocity distribution based on synthetic observations of the two colliding clouds by \citet{tak14}.
%Figure 8c is consistent with the model calculations in colliding clouds in \textcolor{red}{Figure 8d}, lending support for the cloud-cloud collision scenario in G018.149-00.283; the largest velocity in Figure 8d corresponds to that of the small cloud and the gas between the two clouds shows a nearly uniform velocity gradient.
%A difference is that the cavity created by the small cloud is not so clear in the present observation, possibly because of the small extension of the large cloud to the west.

\section{Discussion}
We presented the molecular gas distribution in the RCW 166 region based on the JCMT archival data. The region is divided into three parts of H{\sc ii} bubbles, G018.149-00.283, N21, and N22. In the following we test a cloud-cloud collision scenario by considering the cloud-cloud collision model for RCW 120 by \citet{tor15}. In the scenario a small cloud and a large cloud collide and the small cloud creates a cavity in the large cloud \citep{hab92,ana10,tak14}. 

%移動してきた%
\citet{tak14} made hydrodynamical numerical simulations of two colliding spherical clouds. One of the clouds, the small cloud, has a radius of 3.5 pc an the other, the large cloud has a radius of 7.2 pc. Details of the head-on collision are summarized in Table 2. Figure 6a shows a schematic view of the collision seen from the direction perpendicular to the cloud relative motion at two epochs, 0.0 Myr and 1.6 Myr, after the onset of the collision based on the numerical simulations. The small cloud is producing a cavity in the large cloud by the collisional interaction at 1.6 Myr. The interface layer of the two clouds has enhanced density by collision, and the internal turbulence mixes the inhomogeneous gas of the clouds into the layer. Figures 6b-6i show velocity-channel distributions every 0.9 km s$^{-1}$ as seen from a direction $\theta$ = $45^{\circ}$ tilted to the relative motion.
The small cloud is seen at a velocity range from 2.0 to 4.7 km s$^{-1}$, and the large cloud from -1.5 to 1.1 km s$^{-1}$.
The velocity range from 1.1 to 2.0 km s$^{-1}$  corresponds to the intermediate velocity layer.
Note that the velocity ranges of each panel in Figures 6b-6i do not exactly fit the velocity ranges of the two clouds in part due to the internal turbulent.
The small cloud is flattened perpendicular to the traveling direction by the merging into the intermediate layer, and the large cloud shows intensity depression corresponding to the cavity created by the small cloud.
%%%

In this situation, consequently, the collision compresses gas in the two clouds into an interface layer, which becomes gravitationally unstable, leading to formation of high-mass star(s) inside the cavity. 
This provides an alternative to the usual stellar wind/ionization bubble scenario discussed in the literature (e.g., Zavagno et al. 2010). %\citep{2010A&A...518L..81Z}
An important difference between the two scenarios is that the collision scenario explains the formation of the first generation star as caused by collision, whereas the wind-bubble model has no explanation on the first star formation.
It remains to be tested if the second generation stars are formed by the wind compression or by the collision at compression.
A recent estimate of a cloud-cloud collision frequency in hydrodynamic simulations of isolated galaxies suggests that collision between molecular clouds is frequent enough to happen, every several Myr, within a lifetime of a giant molecular cloud 20 Myrs \citep{fuj14, dob15}. Therefore, we need to explore if the collision is a viable scenario in RCW 166. 
   
A schematic of the model is shown in Figure 8. This indicates that the small cloud produces a cavity in the large cloud and the interface between the two clouds forms a dense shock-compressed interface layer which leads to star formation. In an early phase of the collision, only two clouds are seen without stars and this is phase I which is difficult to recognize because there is no obvious sign of high-mass star formation. In the next step, phase II, an O or B star forms in the interface where an H{\sc ii} region is not yet highly developed. This is a phase where a compact H{\sc ii} region without extended ionization is observed and the two colliding clouds with different velocity are observable as complementary distribution. Then, phase III with extensive ionization will follow, where the inside of the cavity is filled with H{\sc ii} gas ionized by an O star and the small cloud is dissipated by ionization and collisional merging into the compressed layer. Remnant gas of the small cloud may remain outside the ring, depending on the initial cloud morphology; a case in RCW 120 shows such a remnant cloud just outside the bubble opening \citep{tor15}.

\subsection{Possibility of cloud-cloud collision in G018.149-00.283, N21, and N22}
G018.149-00.283 showing the most intense CO emission seems to be the youngest high-mass star formation in RCW 166. 
%移動してきた%
Figure 7d shows simulations of the cloud velocity distribution based on synthetic observations of the two colliding clouds by \citet{tak14}.
Figure 7c is consistent with the model calculations in colliding clouds in Figure 7d, lending support for the cloud-cloud collision scenario in G018.149-00.283; the largest velocity in Figure 7d corresponds to that of the small cloud and the gas between the two clouds shows a nearly uniform velocity gradient.
A difference is that the cavity created by the small cloud is not so clear in the present observation, possibly because of the small extension of the large cloud to the west.
%%%
Cloud-cloud collision in phase II is a possible scenario to be applied to G018.149-00.283; the small cloud with an H{\sc ii} region which is surrounded by a molecular cavity is explained as in Figure 8 (phase II). 
In the present observations we find two clouds at 54 km s$^{-1}$ and 50 km s$^{-1}$, where the small cloud is red-shifted, indicating that the small cloud is moving away from us. 
The H{\sc ii} region is shifted from the compressed layer because the small cloud is moving quickly. 
The two clouds are linked continuously in velocity (Figure 7c) and show complementary distribution (Figure 5c), demonstrating the typical observational signature of cloud-cloud collision.
The collision parameters are as follows; the relative velocity is 4 km s$^{-1}$ for a depth of the bubble of 6 pc in projection.
A ratio between them yields roughly $\sim$ 1.5 Myr for the collision time scale. 
The column density and molecular gas may be uncertain due to ionization by the formed O star. 
By assuming that the ionization is insignificant, we estimated the masses of the small cloud and the ring cloud to be 1.7 $\times$ 10$^4$ $M_{\odot}$ and 1.4 $\times$ 10$^4$ $M_{\odot}$, respectively if the LTE is adopted for GRS data $^{13}$CO($J$=1--0).

In N21, the collision scenario is also applicable to explain the star formation, because there are two clouds, one at 45 km s$^{-1}$, a candidate for the large cloud (ring cloud), follows the bubble shape and the other at 48 km s$^{-1}$ inside the bubble, a candidate for the small cloud. 
We estimated the masses of the ring cloud (45 km s$^{-1}$) and the small cloud (48 km s$^{-1}$) to be 0.8 $\times$ 10$^4$ $M_{\odot}$ and 0.9 $\times$ 10$^4$ $M_{\odot}$, respectively, by assuming the LTE from GRS data $^{13}$CO($J$=1--0) with excitation temperature $T_{\rm ex}=20$ K \citep{and09b} and abundance ratio $X_{\rm ^{13}CO/H_2}=2.0\times10^{-6}$ (e.g., \cite{sim01}).
We remark that the collision scenario is appropriate in N21 since the molecular gas shows no apparent sign of expansion (Figure 3), which is similar to RCW 120.
This indicates that stellar feedback is not effective to form the ring.
One O star is located in the bubble and they are able to ionize inside of the ring. N21 would correspond to phase III, where the small cloud is not yet fully dissipated. The small relative velocity may be due to a projection effect, and the collision time scale is crudely estimated to be $5$ pc $/$ $5$ km s$^{-1}= 1$ Myr. 

In N22, the number of OB stars and O-star candidates is large, 2 and 11, indicating that the initial column density of the colliding gas was possibly high like 10$^{23}$ cm$^{-2}$ (cf., \cite{fuk18}). 
We estimate the mass of the ring cloud to be $\sim$0.5 $\times$ 10$^4$ $M_{\odot}$ by assuming the LTE from GRS data $^{13}$CO($J$=1--0).
There is one velocity component at 50 km s$^{-1}$ which delineates the infrared ring, while we see no observed hint for a secondary cloud inside the ring.
We suggest that the 50 km s$^{-1}$ cloud extending to the west is possibly a remnant of the secondary cloud, which is analogous to RCW 120 having a remnant of the other cloud located outside the ring. 
N22 would correspond to a late stage phase III, where the ring molecular gas is becoming dissipated by ionization. 
The dynamical age of the H{\sc ii} region in N22 was estimated to be 0.06--0.15 Myr by \citet{ji12}, although the ambiguity is large. 
%The time scale of the O-star candidates is roughly estimated to be 1 Myr. 
It is possible that the evolution of N22 have been accelerated by the large number of O stars. 
%The large number of O stars may have accelerated the evolution. 

\subsection{Star formation in RCW 166}
An overall scenario in RCW 166 is as follows. 
The initial state is a large central cloud associated with G018.149-00.283, N21, and N22 having total mass of $\sim 1.0 \times 10^5$ $M_{\odot}$. 
Three small clouds with different velocity collided with the central cloud in the last Myr. 
We can identify candidates for these small clouds in Figure 4b; for G018.149-00.283, the inner red-shifted cloud (Figure 7b), for N21, the inner intermediate velocity cloud (Figure 4b), and for N22, the outer blue-shifted cloud in the west of N22 (Figure 4c) or the outer intermediate velocity cloud in the north of N22 (Figure 4b). 
It is possible that lower-mass YSOs formed without trigger are distributed where collision is not taking place (circles in Figure 1a). 
This scenario suggests that three collisions created the three bubbles, and that the RCW 166 region includes three bubbles of different evolutionary stages in the order [Phase II, G018.149-00.283], [early phase III, N21] and [late phase III, N22].

The present case suggests a fairly high frequency of cloud-cloud collision. The region is located at 4 kpc from the sun and the radius from the Galactic center is about 4 kpc. This indicates a site inside the molecular ring where cloud-cloud collision is frequent. The high cloud density in the molecular ring suggests a collision mean free time to be short, several Myrs \citep{fuj14,dob15}. %
 In a localized region where cloud density is even higher than the average the timescale is proportional to (cloud number density)$^{-1}$ and, can be shorter. This may explain the fairly high frequency of collision in the RCW 166 region. The long filamentary cloud distribution may also favor such frequent collisions like in case of the filamentary clouds in N159 in the LMC \citep{fuk15,sai17}.

The above argument is a unique scenario for O star formation. It is possible that the O stars and H{\sc ii} regions were formed by cloud-cloud collision in N21 and N22. It is however worth mentioning that there are $\sim$ 30 low-mass YSOs around the present bubbles, and they are all distributed outside the bubbles, whereas the O stars are all located inside the bubbles which are suggested to be formed by cloud-cloud collision (Figures 1 and 4 ). This favors an interpretation that O stars are preferentially formed under external trigger, most probably by cloud-cloud collision, and low-mass stars formed without collisional triggering. Apparently, the number of O/B early stars is large as compared with that of observed low-mass stars, suggesting that the region may have a top-heavy stellar mass function as compared with the field stellar mass function with a Salpeter-like slope. This does not strongly affect the universal IMF in general because the number of O stars formed by triggering is yet very small and the evolution is very rapid for the O stars.

%\textcolor{red}{Finally, we discuss G018.305-00.391. By assuming a distance 13.3 kpc to G018.305-00.391, it is possible that this is another H{\sc ii} region formed by triggering in cloud-cloud collision as indicated by the reasonably good complementary distribution between the two velocity components; Figure 5c shows that the peak coincides well with the depression if a displacement is considered between them. The collision timescale is estimated to be $3$ pc $/$ $12$ km s$^{-1}$ $\sim$ a few times 10$^5$ yrs, and is consistent with stage II. We will present fuller account of this H{\sc ii} region separately.}

\section{Conclusions}
We have carried out a new analysis of molecular gas toward the RCW 166 region by using the JCMT CO ($J$=3--2) archival data. We found molecular gas associated with three bubbles and interpreted the results in terms of the cloud-cloud collision model presented for RCW 120 by \citet{tor15}. The main conclusions are as follows;

\begin{enumerate}
\item The $^{12}$CO($J$=3--2) data show the molecular gas at 0.3 pc resolution associated with the three H{\sc ii} regions including two Spitzer bubbles N21 and N22, and the H{\sc ii} region G018.149-00.283. The molecular gas shows ring-like distributions similar to these bubbles.
 The total mass of the molecular gas associated with G018.149-00.283, N21, and N22 is $\sim 1.0 \times$ 10$^5$ $M_{\odot}$ within 20 pc.

\item We found that in G018.149-00.283 two clouds are possibly colliding with each other to trigger formation of an O star which is ionizing an H{\sc ii} region that is heavily obscured. The two clouds show complementary distribution and the intermediate velocity feature, suggesting an interface layer formed by the collision. G018.149-00.283 may represent an evolutionary stage earlier than RCW 120 where the cloud inside the ring is not yet fully ionized, and offers a promising site to pursue the effect of on-going collision in the molecular gas distribution.

\item Bubbles N21 has molecular gas associated with the infrared ring and a secondary gas component which is associated with the young O star(s) inside the ring. The inner cloud is significantly smaller than  the inside of the ring and does not show complementary distribution. This may be due to the ionization of the inner cloud by the four O stars.  

\item N22 is another bubble with ring-like molecular gas. We do not see molecular gas inside the bubble, whereas there are candidates for a secondary cloud just outside the ring. N22 has one O- and one B-type star and 11 O-star candidates inside the ring and dominate high-mass formation in RCW 166. The small cloud which possibly collided here may be already dissipated by collisional merging and ionization.

\item The molecular gas is rich with filamentary distributions in RCW 166. The average mean free time between GMC collision is estimated to be 7 Myrs at R = 6 kpc (e.g., Fujimoto et al. 2014; Dobbs et al 2015). The collision interval 1 Myr is smaller than the timescale. The RCW 166 region is more enhanced in cloud density as manifested many filamentary clouds, which collide possibly more often than the Galactic average.
\end{enumerate}

The evolution of collision between small and large clouds is traced in the three bubbles. In RCW 120 the inner cloud is already dissipated, whereas in G018.149-00.283 the inner cloud, the shocked interface layer, is still observable and shows complementary distribution with the cavity cloud as expected by the cloud-cloud collision model. This indicates G018.149-00.283 is in a stage earlier than RCW120. We suggest that the RCW 166 region includes three bubbles of different evolutionary stages in the order [Phase II, G018.149-00.283], [early phase III, N21] and [late phase III, N22]. 
%Another H{\sc ii} region G018.305-00.391 in the field, perhaps being located by chance, shows that a similar collision may be operating elsewhere within the solar circle.

\begin{ack}
This work was financially supported Grants-in-Aid for Scientific Research (KAKENHI) of the Japanese society for the Promotion for Science (JSPS; grant number society for 15K17607 and 15H05694). 
The James Clerk Maxwell Telescope (JCMT) is operated by the East Asian Observatory on behalf of The National Astronomical Observatory of Japan, Academia Sinica Institute of Astronomy and Astrophysics, the Korea Astronomy and Space Science Institute, the National Astronomical Observatories of China and the Chinese Academy of Sciences (Grant No. XDB09000000), with additional funding support from the Science and Technology Facilities Council of the United Kingdom and participating universities in the United Kingdom and Canada.
This publication also makes use of molecular line data from the Boston University-FCRAO Galactic Ring Survey (GRS). The GRS is a joint project of Boston University and Five College Radio Astronomy Observatory, funded by the National Science Foundation under grants AST-9800334, AST-0098562, AST-0100793, AST-0228993, and AST-0507657.
In addition, this work is based in part on observations made with the Spitzer Space Telescope, which is operated by the Jet Propulsion Laboratory, California Institute of Technology under a contract with NASA.
\end{ack}

%%%
% See the manual for the detail.
%%%

%\bibliographystyle{apj}
%\bibliography{reference_rcw166}

\newpage
\begin{table*}
%\begin{center}
\tbl{Molecular clouds in RCW 166}{% 
\begin{tabular}{ccccccccc} 
\hline
source & $l$ & $b$ & Distance (kpc) & Velocity (km s$^{-1}$) & Mass of molecular clouds$^{\dag}$ ($M_{\odot}$) & Maximum $N({\rm H_2})$$^{\dag}$ (cm$^{-2}$)\\
\hline
\hline
G018.149-0.283 (blue)\footnotemark[1] & 18\fdg15 & -0\fdg28 & 4.0 & 46.6 -- 49.6 & $1.7\times 10^{4}$ & $1.4\times 10^{22}$ \\
-- (intermediate)\footnotemark[2] & -- & -- & -- & 49.6 -- 51.6 & $1.2\times 10^{4}$ & $1.4\times 10^{22}$ \\
-- (red)\footnotemark[3] & -- & -- & -- & 51.6 -- 54.6 & $1.4\times 10^{4}$ & $2.2\times 10^{22}$ \\
%G018.237-0.240 & 18\fdg237 & -0\fdg240 & 4.0 &  &  & \\
%G018.305-0.391 (blue)\footnotemark[4]  & 18\fdg31 & -0\fdg39 & 13.3  & 26.6 -- 36.6 & $6.8\times 10^{4}$ & $2.2\times 10^{22}$ \\
%-- (red)\footnotemark[5]  & -- & -- & -- & 40.6 -- 48.6 & $8.6\times 10^{4}$ & $1.2\times 10^{22}$ \\
N21 (ring)\footnotemark[4] & 18\fdg19 & -0\fdg40 & 4.0  & 43.6 -- 47.6 & $0.8\times 10^{4}$ & $1.4\times 10^{22}$ \\
-- (red)\footnotemark[5] & -- & -- & -- & 47.6 -- 50.6 & $0.9\times 10^{4}$ & $1.2\times 10^{22}$ \\
N22 & 18\fdg25 & -0\fdg31 & 4.0  & 47.6 -- 50.6 & $0.5\times 10^{4}$ & $0.8\times 10^{22}$ \\
\hline
%\noalign{\vskip3pt} 
\end{tabular}} 
\label{tab:first} 
\begin{tabnote}
%{\hbox to 0pt{\parbox{150mm}{\footnotesize
Note. 
\footnotemark[$\dag$] $N({\rm H_2})$ and cloud mass were derived from GRS $^{13}$CO($J$=1--0) data by assuming LTE with $T_{\rm ex}=20$ K and $X_{\rm ^{13}CO/H_2}=2.0\times10^{-6}$. \\
\footnotemark[1] G018.149-0.283 (blue) indicates blue-shifted cloud shown Figure 4a. 
\footnotemark[2] G018.149-0.283 (intermediate) indicates intermediate cloud shown Figure 4b.\\
\footnotemark[3] G018.149-0.283 (Red) indicates red-shifted cloud shown Figure 4c. 
%\footnotemark[4] G018.305-0.391 (blue) indicates blue-shifted cloud shown Figure 5a.\\
%\footnotemark[5] G018.305-0.391 (red) indicates red-shifted cloud shown Figure 5b. 
\footnotemark[4] N21 (ring) indicates blue-shifted cloud shown Figure 4a.\\
\footnotemark[5] N21 (red) indicates red-shifted cloud shown Figure 4b.\\
%\par\noindent
%\phantom{0}
%\par
%\hangindent6pt\noindent
%}\hss}}
\end{tabnote} 
%\end{center} 
\end{table*}

\newpage
\begin{table}[h] 
\begin{center}
\tbl{The initial conditions of the numerical simulations \citep{tak14} }{% 
\begin{tabular}{@{}cccccc@{}} \noalign{\vskip3pt}
\hline\hline
\multicolumn{1}{c}{Box size [pc]} & $30 \times 30 \times 30$  &  &  &  \\ [2pt]
\noalign{\vskip3pt}
Resolution [pc] & 0.06 &  &  \\
Collision velocity [km s$^{-1}$] & 10 (7)$^{\dag}$ & & \\
\hline
Parameter & The small cloud & The large cloud & note    \\
\hline
Temperature [K] & 120 & 240 &  \\
Free-fall time [Myr] & 5.31 & 7.29 &  \\
Radius [pc] & 3.5 & 7.2 &  \\
Mass [$M_{\odot}$] & 417 & 1635 &  \\
Velocity dispersion [km s$^{-1}$] & 1.25 & 1.71 &  \\
Average Density [cm $^{-3}$] & 47.4 & 25.3 & Assumed a Bonner-Ebert sphere  \\
\hline\noalign{\vskip3pt} 
\end{tabular}} 
\label{tab:first} 
\begin{tabnote}
{\hbox to 0pt{\parbox{150mm}{\footnotesize
Note. \footnotemark[$\dag$] The initial relative velocity between the two clouds is set to 10 km s$^{-1}$, whereas the collisional interaction decelerates the relative velocity to about 7 km s$^{-1}$ in 1.6 Myrs after the onset of the collision. The present synthetic observations are made for a relative velocity 7 km s$^{-1}$ at 1.6 Myrs.
\par\noindent
%\footnotemark  \par\noindent
%\footnotemark
\phantom{0}
\par
\hangindent6pt\noindent
%\hbox to6pt{\,\footnotemark \hss}\unskip%
}\hss}}
\end{tabnote} 
\end{center} 
\end{table}

\newpage
\begin{figure*}%1
\begin{center}
 \includegraphics[width=15cm]{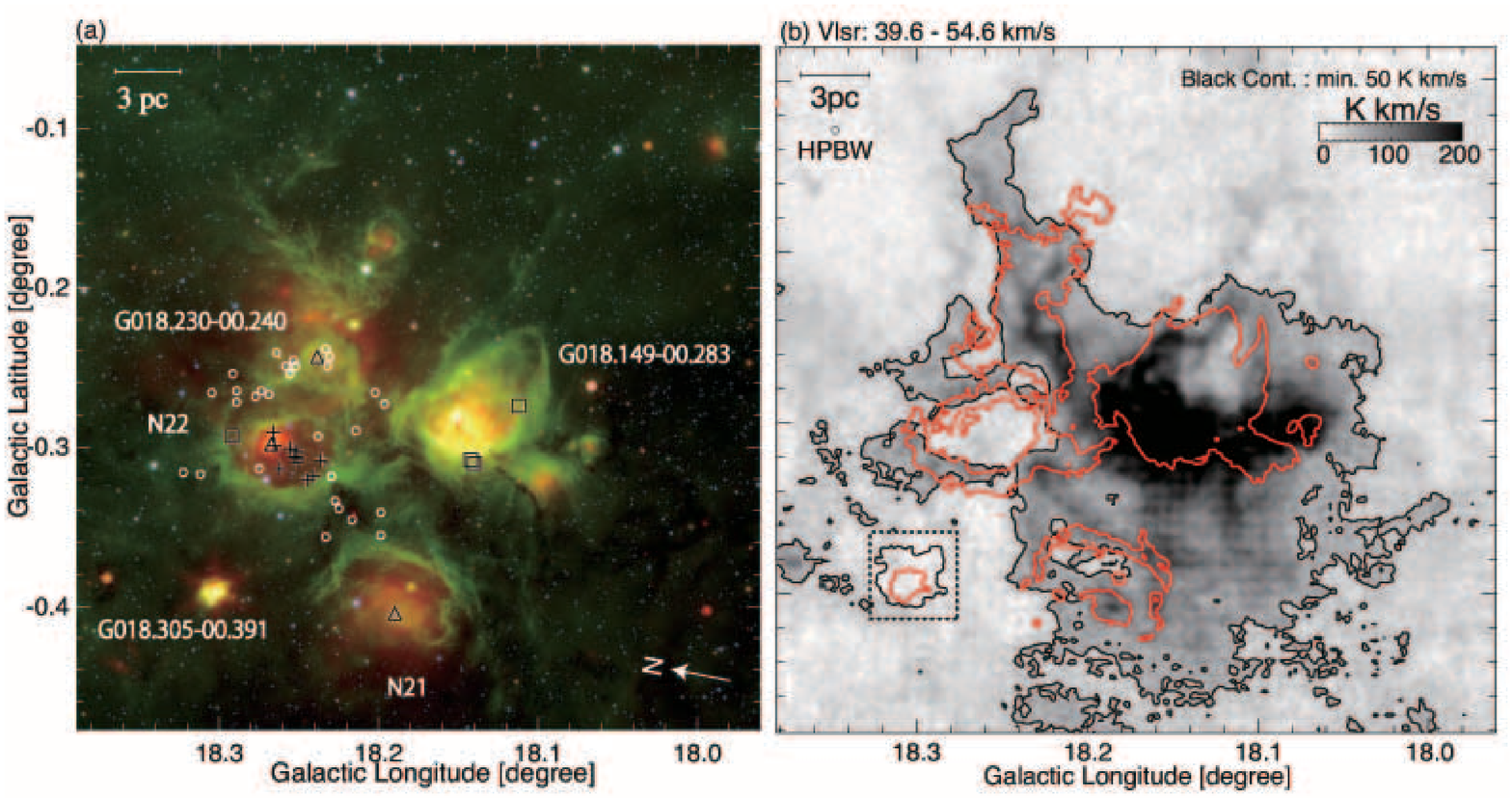}
   \end{center}
\caption{(a) Color composite image of RCW 166. Blue, green and red show the $Spitzer$/IRAC 3.6 $\mu$m, the $Spitzer$/IRAC 8 $\mu$m and $Spitzer$/MIPS 24 $\mu$m. Triangles and squares represent O stars and B stars \citep{par13}, respectively. Crosses and circles indicate O-star candidates \citep{ji12} and YSOs \citep{wat08}, respectively. (b) $^{12}$CO($J$ =3--2) distributions toward RCW166 with the JCMT data set. The image and black contour show the $^{12}$CO($J$ =3--2) emission integrated from 39.6 to 54.6 km s$^{-1}$. The black contour level is from 50 K km s$^{-1}$. The red contours show the outline of the 8 $\mu$m ring. Contours inside the dotted lines the distribution of $^{12}$CO($J$ =3--2) integrated the velocity from 26.6 km s$^{-1}$ to 36.6 km s$^{-1}$ toward the H{\sc ii} region G018.305-00.391. }
\end{figure*}

\begin{figure*}%2
 \begin{center}
  \includegraphics[width=17cm]{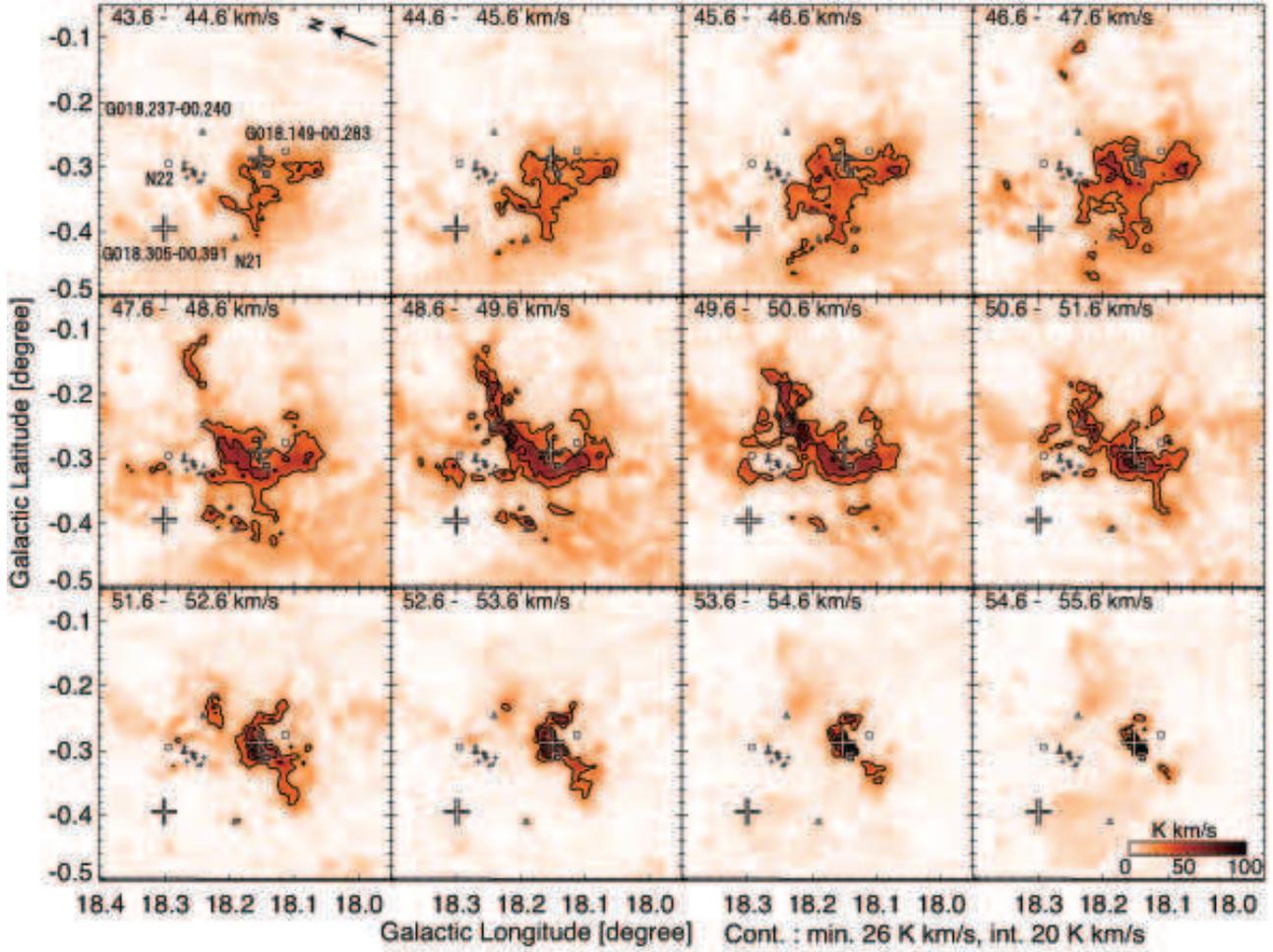} 
 \end{center}
\caption{Velocity channel distributions of the $^{12}$CO($J$=3--2) emission at velocity step of 1 km s$^{-1}$. The image and contour show the distribution of $^{12}$CO($J$=3-2) toward RCW166. The contour level is from 26 K km s$^{-1}$ every 20 K km s$^{-1}$. Triangles and squares represent O stars and B stars \citep{par13}, respectively. Large and small crosses indicate H{\sc ii} regions (G018.149-00.283 and G018.305-0.391) \citep{urq14} and O-star candidates \citep{ji12}, respectively. }
\end{figure*}

\begin{figure*}%3
 \begin{center}
  \includegraphics[width=8cm]{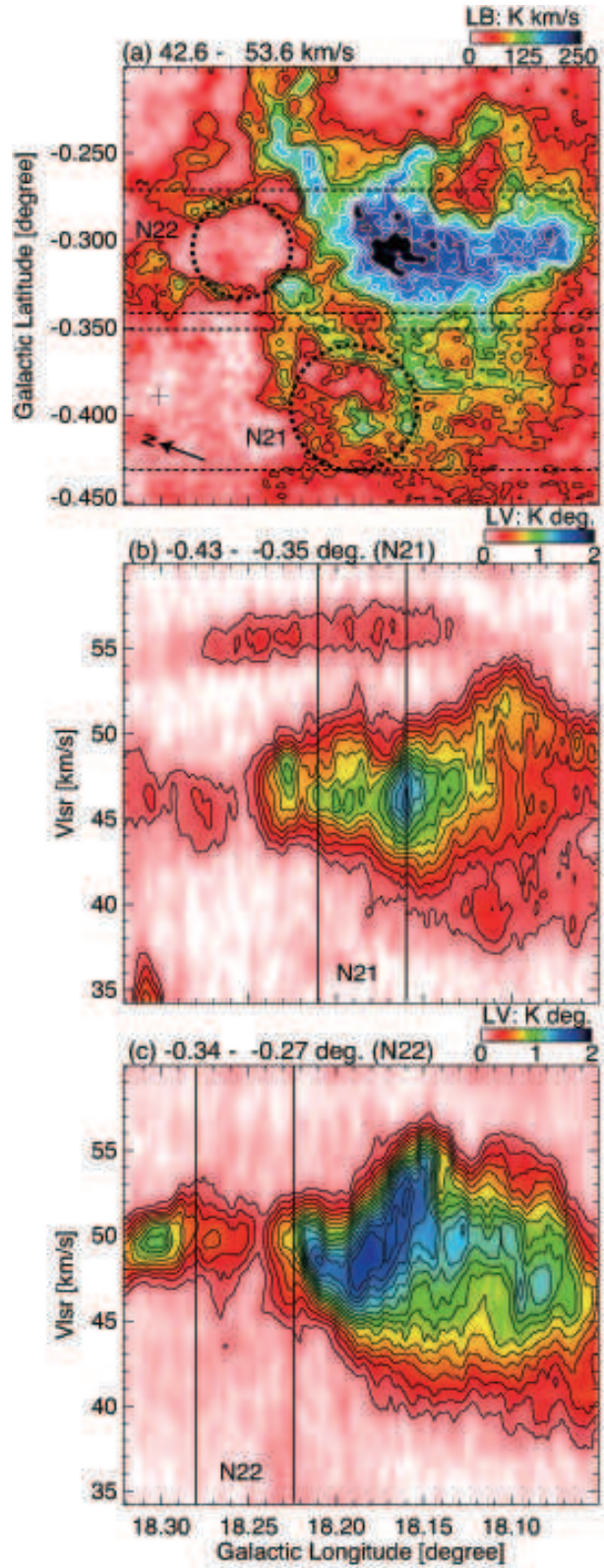} 
 \end{center}
\caption{The image shows two longitude velocity-diagrams toward the bubbles N21 and N22, each of which are marked by dashed circles. a)The $^{12}$CO $J$= 3--2 integrated intensity distribution is shown from 42.6 km s$^{-1}$ to 53.6 km s$^{-1}$. b) Longitude-velocity diagram including N21 and c) that including N22.}
\end{figure*}

\begin{figure*}%4
 \begin{center}
  \includegraphics[width=17cm]{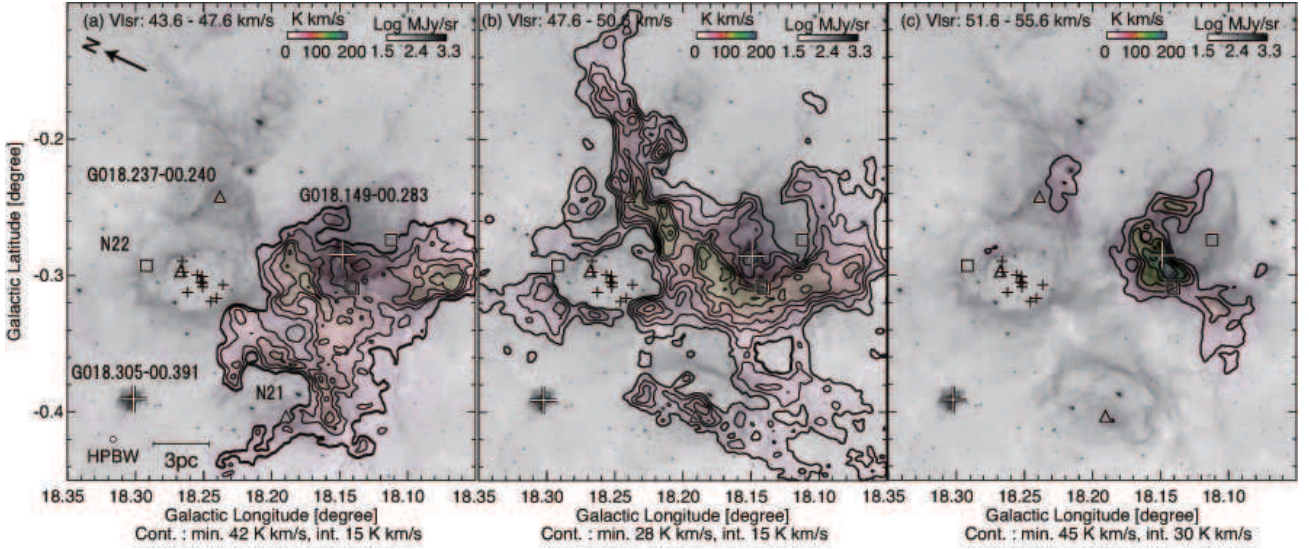} 
 \end{center}
\caption{Comparisons of $^{12}$CO($J$=3--2) distribution (color and contours) of the blue-shifted cloud (a), the intermediate cloud (b) and the red-shifted cloud (c) with the $Spitzer$ 8 $\mu$m emission (grayscale). Triangles and squares represent O stars and B stars \citep{par13}, respectively. Large and small crosses indicate H{\sc ii} regions (G018.149-00.283 and G018.305-0.391) \citep{urq14} and O-star candidates \citep{ji12}, respectively. Panel (a) shows distribution of the ring cloud toward N21. The contour level is from 42 K km s$^{-1}$ every 15 K km s$^{-1}$. Panel (b) shows the distribution of the ring cloud toward G018.149-00.283, the small cloud toward N21 and the ring cloud toward N22. The contour level is from 28 K km s$^{-1}$ every 15 K km s$^{-1}$. Panel (c) shows distribution of the small cloud toward G018.149-00.283. The contour level is from 45 K km s$^{-1}$ every 15 K km s$^{-1}$.}
\end{figure*}

%\begin{figure*}%5
% \begin{center}
%  \includegraphics[width=17cm]{fig/fig6.eps} 
% \end{center}
%\caption{The region of the complementary distribution in the box with solid lines. The cross indicate the H{\sc ii} region G018.305-00.391.
% A distance of 13.3 kpc is assumed (see the text).
% (a) The image and the blue contours show the blue-shifted cloud (26.6 - 36.6 km s$^{-1}$). (b) The image and the red contours show the red-shifted cloud (40.6 - 46.6 km s$^{-1}$). }
 %(c) The image indicates the complementary distribution of the two cloud with the different velocity after the displacement 3 pc shown the arrow of $60^{\circ}$ angle. The dotted and solid lines show the distribution before the displacement and after the displacement. The blue and red contours are from 30 K km s$^{-1}$ every 30 K km s$^{-1}$, respectively.}
%\end{figure*}

\begin{figure*}%6
 \begin{center}
  \includegraphics[width=17cm]{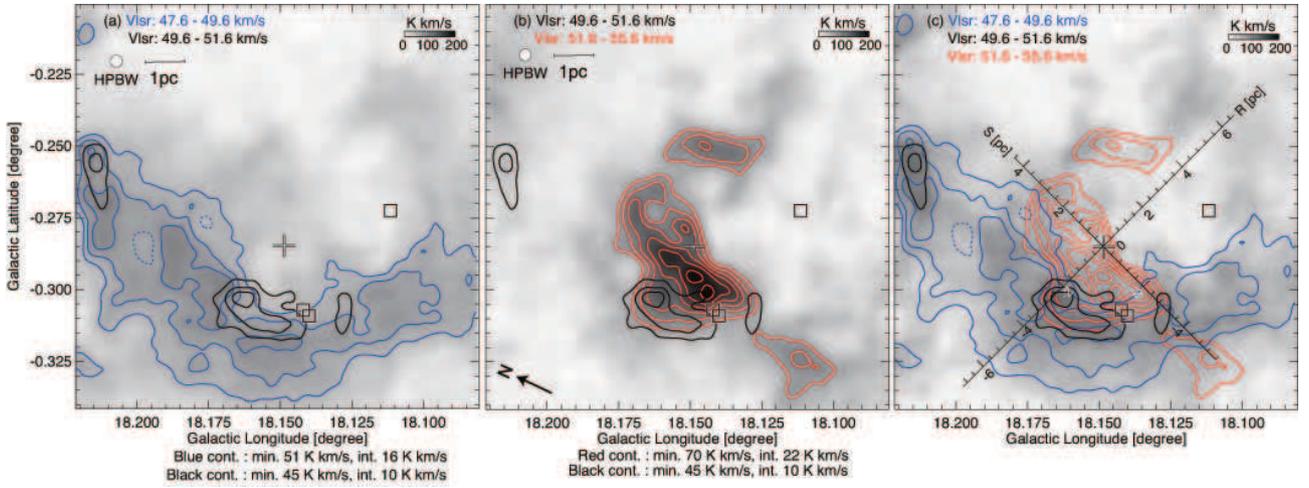} 
 \end{center}
\caption{Molecular gas distribution toward the H{\sc ii} region G018.149-00.283 shown by the cross. (a) The image and the blue contours indicate the blue-shifted cloud (47.6 - 49.6 km s$^{-1}$), and the black contours show the intermediate cloud (49.6 - 51.6 km s$^{-1}$). The blue and black contour levels are from 51 K km s$^{-1}$ every 16 K km s$^{-1}$ and 45 K km s$^{-1}$ every 10 K km s$^{-1}$. Squares represent B stars \citep{par13}. (b) The image and the red contours show the red-shifted cloud (51.6 - 55.6 km s$^{-1}$). The red contour level is from 70 K km s$^{-1}$ every 10 K km s$^{-1}$. (c) The image and the black indicate intermediate cloud. The image indicated the complementary distributions of the three clouds with the different velocity.
R-axis was taken for the direction perpendicular to the elongation of the red-shifted component and the S-axis is orthogonal to the R-axis.
The origin is the central position of the H{\sc ii} region G018.149-00.283.}
\end{figure*}

\begin{figure*}%7
 \begin{center}
  \includegraphics[width=16cm]{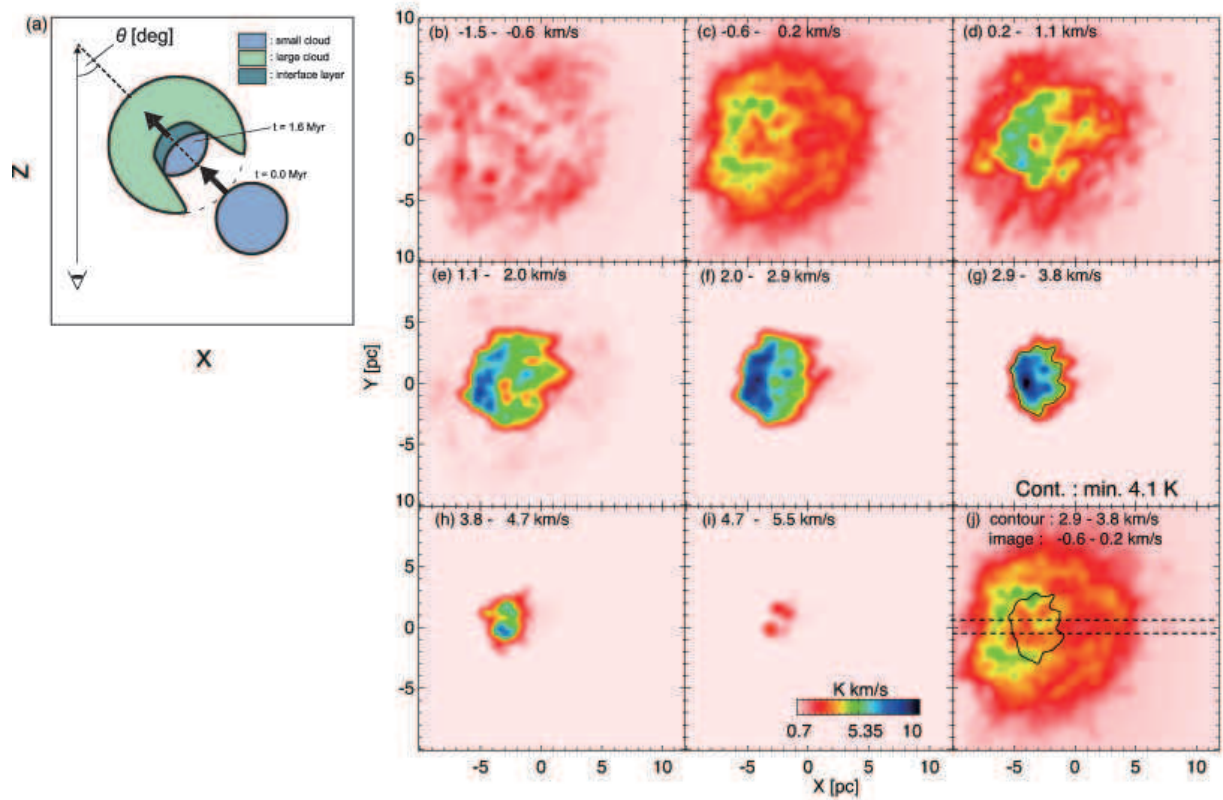} 
 \end{center}
\caption{The numerical model by \citet{tak14} at 1.6 Myr observed at an angle of $45^{\circ}$ between the line of sight and the cloud relative motion. Panel (a) is a schematic of the top-view of the collision. Panels (b-i) show the velocity channel distributions every 0.9 km s$^{-1}$ in a velocity interval indicated in each panel. Panel (j) shows an overlay between the large cloud, the image in Panel (c), and the small cloud with the contour of Panel (g). The black contour is that of 4.1 km s$^{-1}$ in panel (g).}
\end{figure*}

\begin{figure*}%8
 \begin{center}
  \includegraphics[width=7cm]{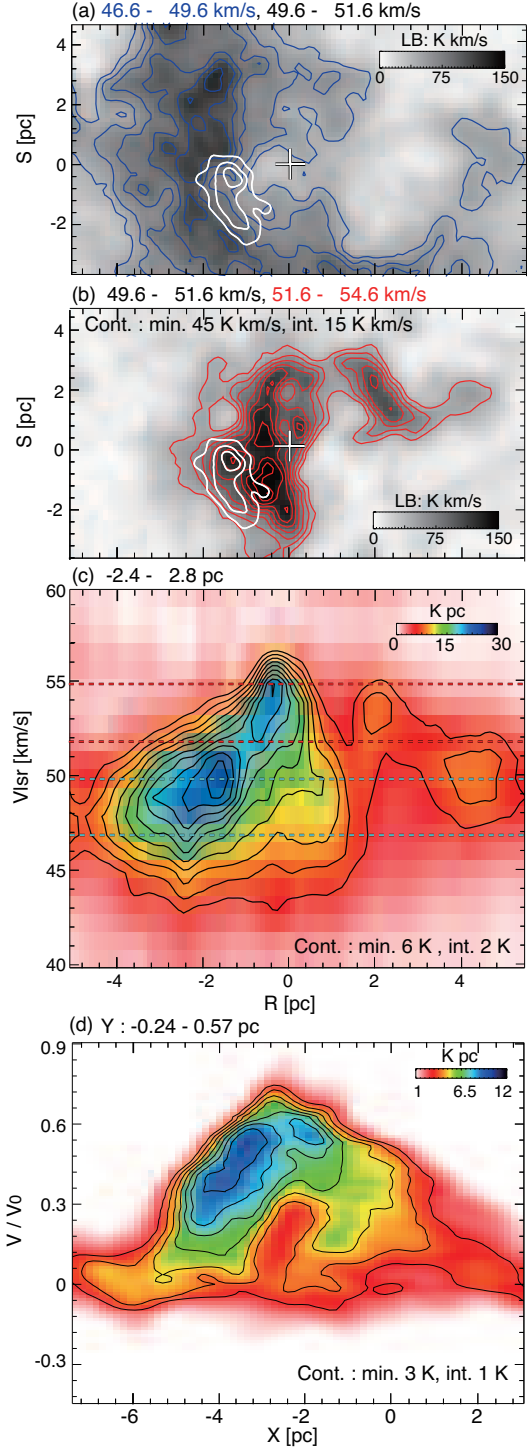} 
 \end{center}
\caption{(a--c) $^{12}$CO($J$=3--2) distribution toward the H{\sc ii} region G018.149-00.283. (a) The image and the white contours show the blue-shifted cloud and the intermediate cloud. The horizontal axis is common with panel (b) and (c), and R-axis is the direction of the displacement of the small cloud as shown Figure 6c. (b) The image and the white contours show the red-shifted cloud and the intermediate cloud. (c) The image and the black contours show the position-velocity diagram to integrate S-axis -2.4 to 2.8 pc. (d) Position-velocity diagram of the synthetic observations by \citet{tak14} same as Figure 7. V and V$_{0}$ indicate the relative velocity between the large and small clouds and the relative velocity 7 km s$^{-1}$ at 1.6 Myrs.}
\end{figure*}

\begin{figure*}%9
 \begin{center}
  \includegraphics[width=5cm]{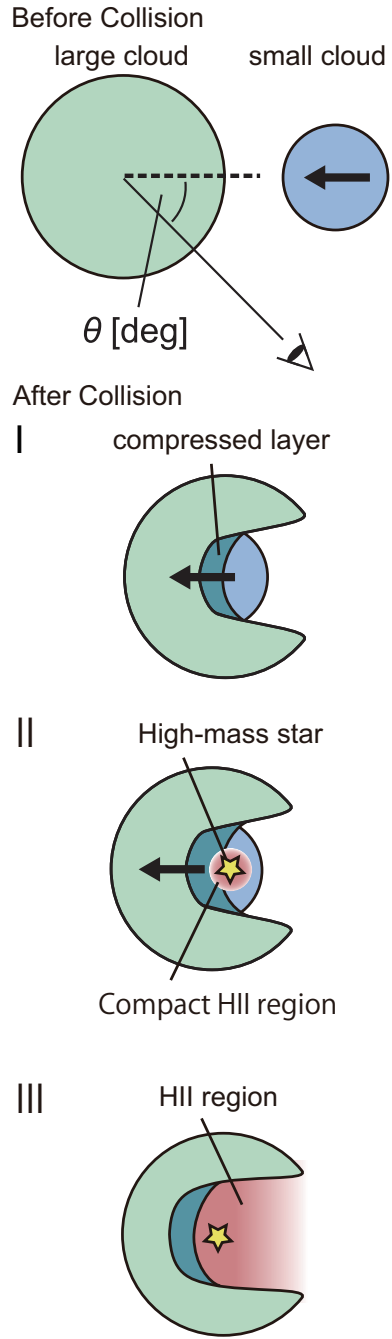} 
 \end{center}
\caption{Schematics of basic scenario of cloud-cloud collision introduced by Habe \& Ohta (1992; see Section 4.2.2 for details). The top image shows a schematic of the top-view of the collision before collision. The green and blue circles are the large and small clouds, respectively. The small cloud is moving to the line of sight as shown the arrow. We observed at an angle of $\theta$ = $45^{\circ}$ between the line of sight and the cloud relative motion. Phase I shows the early stage of the collision that the two clouds are seen without stars. Phase II shows the next stage that O or B star forms in the interface and an H{\sc ii} region is the very small such as the compact H{\sc ii} region. Phase III shows the stage that the inside of the cavity is filled with H{\sc ii} gas ionized by an O star and the small cloud is dissipated by ionization.}
\end{figure*}

\end{document}